%% file: main.tex
\documentclass{article}

\usepackage[final]{neurips_2023_ml4ps}

\usepackage[utf8]{inputenc} 
\usepackage[T1]{fontenc}    
\usepackage{hyperref}       
\usepackage{url}            
\usepackage{booktabs}       
\usepackage{amsfonts}       
\usepackage{nicefrac}       
\usepackage{microtype}      
\usepackage{xcolor}         

\usepackage{subfig}
\usepackage{graphicx}
\usepackage{placeins}
\usepackage{amsmath}

\usepackage[super]{nth}

\usepackage[backend=bibtex,sorting=none,style=numeric-comp]{biblatex}
\bibliography{main.bib}

\title{Smartpixels: Towards on-sensor inference of charged particle track parameters and uncertainties}

\author{Jennet Dickinson$^1$, Rachel Kovach-Fuentes$^2$, Lindsey Gray$^1$, Morris Swartz$^3$,\\ \textbf{Giuseppe Di Guglielmo}$^{1,4}$,
\textbf{Alice Bean}$^5$,
\textbf{Doug Berry}$^1$, 
\textbf{Manuel Blanco Valentin}$^4$,\\
\textbf{Karri DiPetrillo}$^2$,
\textbf{Farah Fahim}$^{1,4}$,
\textbf{James Hirschauer}$^1$,
\textbf{Shruti R. Kulkarni}$^6$, 
\textbf{Ron Lipton}$^1$,\\
\textbf{Petar Maksimovic}$^3$,
\textbf{Corrinne Mills}$^7$,
\textbf{Mark S. Neubauer}$^8$,
\textbf{Benjamin Parpillon}$^{1,7}$,\\
\textbf{Gauri Pradhan}$^1$,
\textbf{Chinar Syal}$^{1}$, 
\textbf{Nhan Tran}$^{1,4}$, 
\textbf{Dahai Wen}$^{3}$,
\textbf{Jieun Yoo}$^7$,
\textbf{Aaron Young}$^6$
\\
\\
$^1$ Fermi National Accelerator Laboratory, Batavia, IL 60510, USA\\
$^2$ The University of Chicago, Chicago, IL 60637, USA\\
$^3$ Johns Hopkins University, Baltimore, MD 21218, USA\\
$^4$ Northwestern University, Evanston, IL 60208, USA\\
$^5$ University of Kansas, Lawrence, KS 66045, USA\\
$^6$ Oak Ridge National Laboratory, Oak Ridge, TN 37831, USA\\
$^7$ University of Illinois Chicago, Chicago, IL, 60607, USA\\
$^8$ University of Illinois Urbana-Champaign, Champaign, IL 61801, USA
}

\begin{document}

\maketitle

\begin{abstract}
The combinatorics of track seeding has long been a computational bottleneck for triggering and offline computing in High Energy Physics (HEP), and remains so for the HL-LHC. Next-generation pixel sensors will be sufficiently fine-grained to determine angular information of the charged particle passing through from pixel-cluster properties. This detector technology immediately improves the situation for offline tracking, but any major improvements in physics reach are unrealized since they are dominated by lowest-level hardware trigger acceptance. We will demonstrate track angle and hit position prediction, including errors, using a mixture density network within a single layer of silicon as well as the progress towards and status of implementing the neural network in hardware on both FPGAs and ASICs.
\end{abstract}

\clearpage

\setcounter{page}{1}

\input{sections/1-introduction/introduction}
\input{sections/2-relatedwork/relatedwork}
\input{sections/3-dataset-model/dataset-model}
\input{sections/4-results/results}
\input{sections/5-conclusions/conclusions}


\printbibliography


\clearpage

\input{sections/6-acknowledgements/acknowledgements}

\end{document}

%% file: sections/1-introduction/introduction.tex
\section{Introduction}

Tracking detectors are an important component of modern particle physics detectors that provide the majority of kinematic information describing the aftermath of a collision of fundamental particles.
The tracking detector's basic function is the measurement of space points, i.e. the ``hits" on a ``track", along the trajectories of charged particles that are typically being deflected by a magnetic field, without significantly altering the particles' trajectories during these measurements.
Those space points are processed later by pattern recognition algorithms to yield instances of track hypotheses, each with estimates of the true parameters of the charged particle's kinematics.
Achieving these measurements often requires algorithms that scale with the number of possible combinations of all input data, and this causes a significant increase in an experiment's computational needs for future colliders such as HL-LHC, FCC-hh, or a muon collider.
Any new technology that restricts the number of combinations improves the physics impact of these future colliders by reducing this compute burden and allowing for more detailed downstream analysis. 
If a large enough reduction is achieved, it makes real-time trigger systems possible that drastically alter the physics reach of detectors.
Focusing on trackers designed using silicon sensors, this reduction is traditionally achieved using special double-layers with local pattern recognition logic to select interesting hit pairs yielding coarse track hypotheses that permit fewer valid combinations when considering the output of other layers.
In this work we demonstrate the feasibility of making single-layer track trajectory measurements using a pixelated silicon device, including credible error estimates, by applying machine-learning based uncertainty prediction techniques in quantized neural networks that can be rendered into both FPGA and radiation-hard circuits on the pixel readout ASIC. 
Furthermore, given the general nature of the techniques used, this technology could be beneficially applied in other real-time, on-device pattern recognition tasks.
This technique improves the information supplied by a single layer of silicon, while reducing the material budget necessary to make such measurements, to the point that a pixel-based trigger system becomes feasible for the challenging environments expected at the HL-LHC and beyond.

%% file: sections/2-relatedwork/relatedwork.tex
\section{Related Work}

Usage of neural networks in pattern recognition problems is commonplace in HEP, though typically as a discriminant or refinement to the output of traditional pattern recognition algorithms.
Both the CMS~\cite{Chatrchyan:1129810, elmetenawee2020cms} and ATLAS~\cite{Aad:1129811,Khoda:2019gm} collaborations employ neural networks in their current reconstructions to improve the processing efficiency by removing or correcting mis-reconstructed space points or trajectory hypotheses.
Expanding on this theme, next-generation pattern recognition algorithms that are based mostly or wholly in neural architectures are merging these refinements with the higher-level pattern recognition~\cite{Fox_2021} to reduce the combinatorial fake rate and corresponding computational load.
Our contribution to this avenue of research is to refine the techniques discovered so far by providing a complete statistical interpretation of the space-point and trajectory angle information, and to synthesize the network predicting these data into a form that is implementable in detector front-end hardware.
This creates the possibility for novel, intelligent, and low mass detectors that can significantly improve triggering capabilities by providing space points close to the interaction region that do not cause egregious combinatorial growth in the number of patterns to consider.

%% file: sections/3-dataset-model/dataset-model.tex
\section{Dataset and model architecture}
\label{sec:dataset-model}

The studies in this paper are based on a simulated dataset of silicon pixel clusters produced by charged particles (pions)~\cite{zenodo}.
The kinematic properties of the incident particles are taken from fitted tracks in CMS 13 TeV proton-proton collision data. To study a concrete sensor configuration, charge deposition is simulated in a 21$\times$13 array of pixels described by coordinates $x\times y$, with the $z$ direction normal to the sensor plane. The position $(x,y)$ where the charged particle traverses the sensor mid-plane is assumed to be uniform across the central $3\times3$ pixel array.  The sensor is taken to be 100\,$\mu m$ thick with pixel pitch 50\,$\mu m$ $\times$ 12.5\,$\mu m$ in $x\times y$. A bias voltage of -100V is applied and the detector is immersed in a 3.8\,T magnetic field parallel to the $x$ coordinate.  The particle origin point is taken to be 30mm from the sensor plane. 
 
\begin{figure}[htbp]
  \centering
  \includegraphics[width=0.8\textwidth]{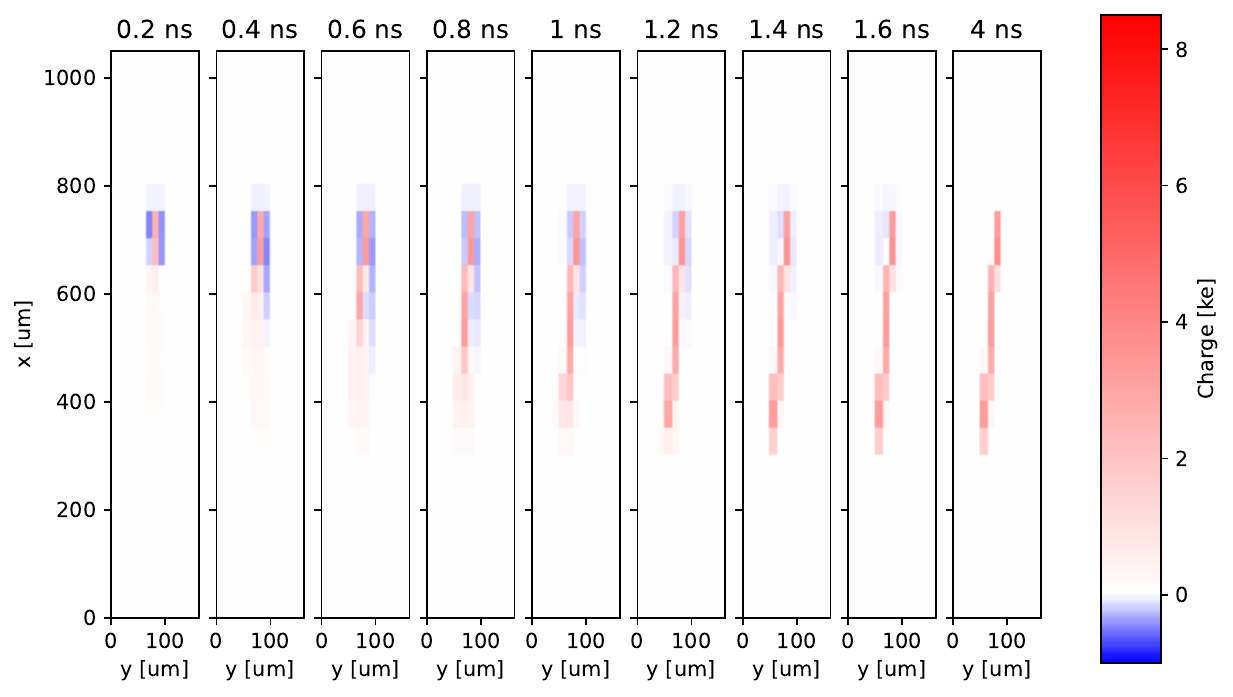}
  \caption{Charge deposition at different times during the cluster evolution for an example cluster. The color scale represents the collected charge in kilo-electrons, with blue representing negative induced charge.}
  \label{fig:cluster}
\end{figure}

The response of this detector is simulated using a time-sliced version of PixelAV~\cite{pixelav}. This provides an accurate model of charge deposition by primary hadronic tracks, a realistic electric field map resulting from the simultaneous solution of Poisson’s Equation, carrier continuity equations, and various charge transport models, an established model of charge drift physics, a simulation of charge trapping and the signal induced from trapped charge, and a simulation of electronic noise, response, and threshold effects.  PixelAV also provides the time evolution of the drift and induced currents in the pixel sensor, and the charge deposition is sampled at 20 time points each separated by 200ps. Figure \ref{fig:cluster} shows the time evolution of the cluster for an example particle, and the training dataset contains 3 million examples of such clusters. 

The shape of the charge deposited in the pixel array is sensitive to the particle's impact position and angle of incidence. The collected charge is log-compressed, scaled to fit in the range $[-1,1]$ and then quantized into 4 bits with a sign bit and 3 bits below the decimal point. The incident angle in the $x-z$ plane is denoted by $\alpha$, and by $\beta$ in the $y-z$ plane (the bending plane of the magnetic field). 

In order to achieve a well-performing network that can reduce combinatorics in downstream reconstruction tasks, we employ a mixture density network (MDN) ~\cite{mdn_paper} with some specializations to the task at hand.
In order to provide a trajectory state estimate at the mid-plane of the pixel sensor the model must predict the local x, y, $\alpha$, and $\beta$ of the cluster, as well as the associated covariance matrix.
The response and resolution distributions are inherently multi-Gaussian, but safely approximated as Gaussian per-cluster, from the varying number of hits per cluster.
Therefore, the network predicts the parameters of a single multi-dimensional Gaussian, and we construct the mixture of the MDN at the level of the likelihood over the training data.  
This likelihood is the loss function that is minimized for this machine learning task.

\begin{figure}[htbp]
  \centering
  \includegraphics[angle=90,width=0.95\textwidth]{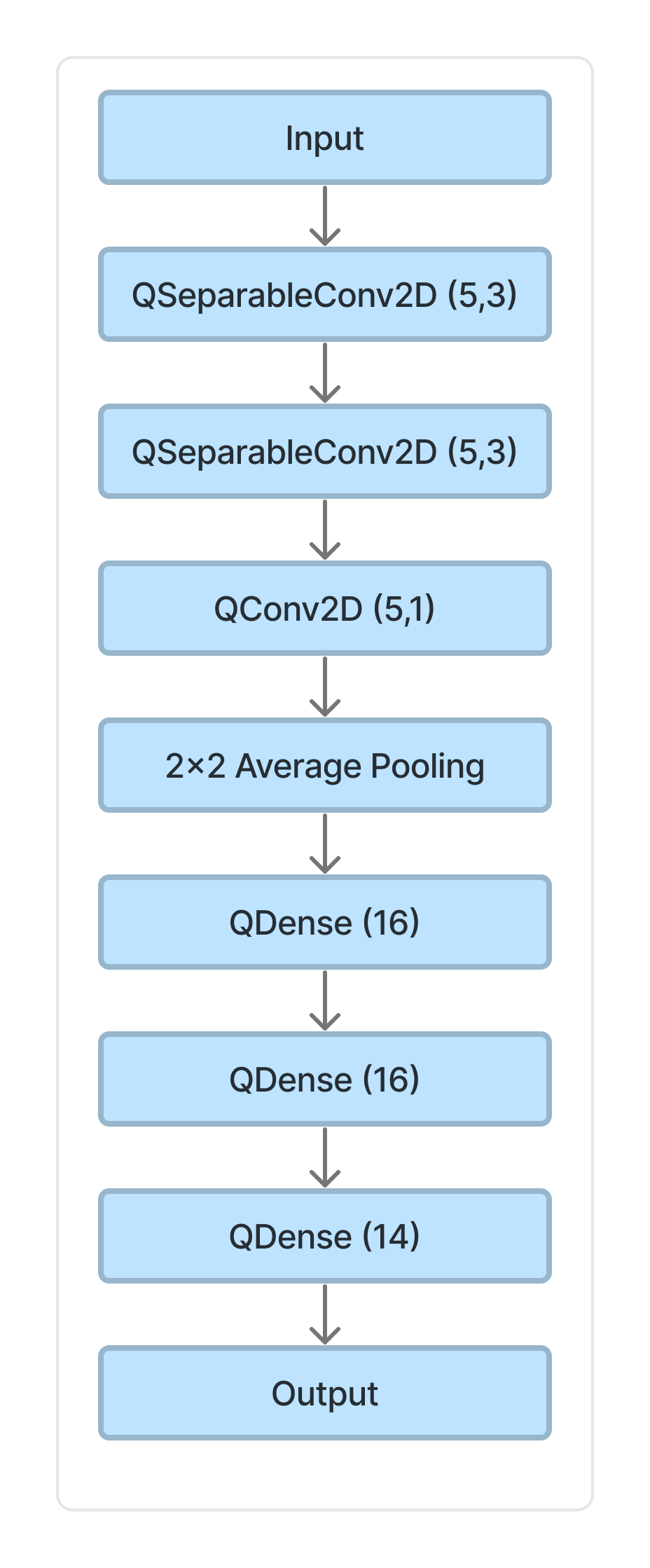}
  \caption{A block diagram of the model architecture used for the mixture-density network based regression of track parameters. The first two separable convolution layers output five filters have have a 3x3 kernel, and the final pointwise convolution outputs 5 filters. The first two dense layers have 16 output units and the last dense layer has 14. The convolutional layers have 4-bit quantized weights with 3 bits below the decimal point and a sign bit, and the dense layers have 8-bit quantized weights with 7 bits below the decimal point and a sign bit. All layers except for the average pooling are activated with a "hard-tanh" function that has the same quantization as for the respective layer weights. The average pooling is 8-bit quantized with 7 bits after the decimal point and a sign bit.}
  \label{fig:networkarch}
\end{figure}

The model itself is a 6 layer quantized network with 3 convolutional layers and 3 dense layers built using the QKeras package \cite{Coelho_2021}, which enables a quantization-aware training. The model is then trained on the previously described dataset for 300 epochs. The architecture of the model is shown in Figure~\ref{fig:networkarch}. The convolutional layers are implemented as depthwise-separable convolutions to reduce the number of model parameters and total operations needed and the averaging pooling layer is introduced to reduce the number of dense-layer parameters by nearly a factor of 4, to minimize on-device resource usage. The optimization of the parameter bitwidths was performed by hand, focusing on reducing the bitwidth in the computationally expensive convolutional layers as much as possible while still retaining good regression performance. This has the particular benefit of keeping the necessary bitwidth for accumulators, registers that contain total results of multiply-accumulate operations, beneath the threshold of requiring complicated multipliers to be used or implemented during synthesis for target devices. To translate the algorithm from a quantized graph representation into an efficient hardware implementation, we use \texttt{hls4ml}, an open-source Python framework for co-design~
\cite{vloncar_2021_5680908,Duarte:2018ite}. 

\FloatBarrier

%% file: sections/4-results/results.tex
\section{Results}

\begin{figure}[!htb]
    \centering
    \subfloat[]{\includegraphics[width=0.45\textwidth]{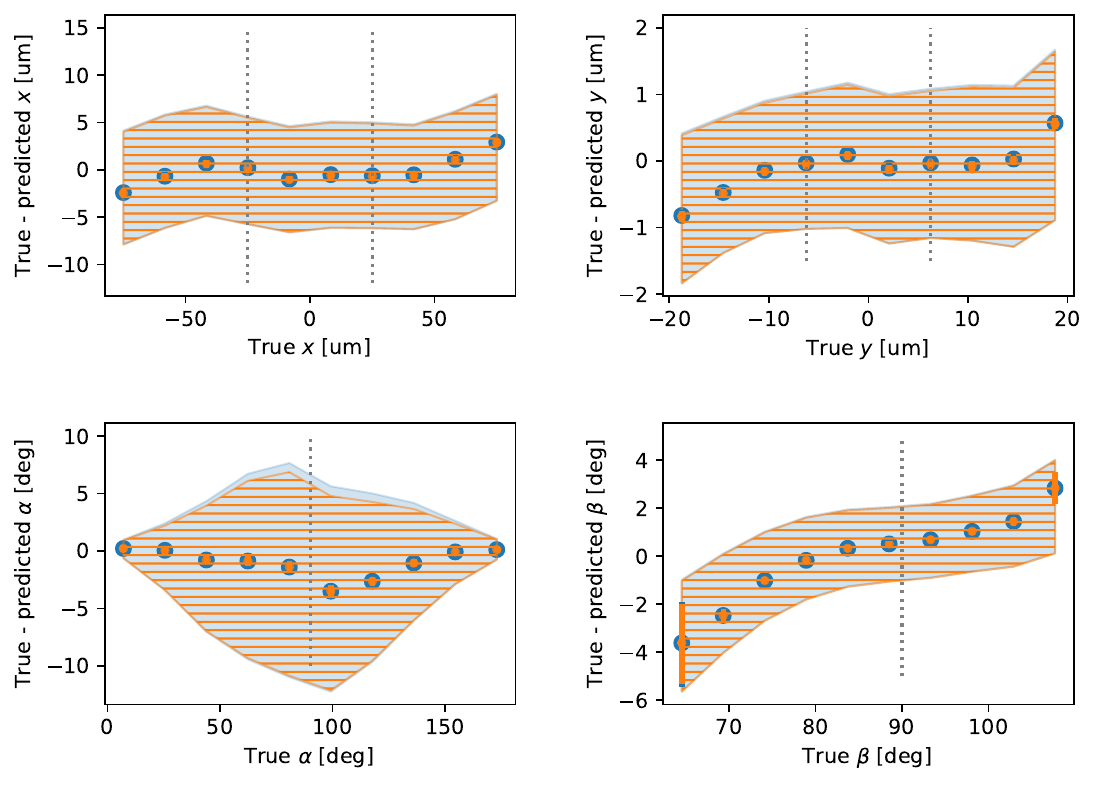}}
    \subfloat[]{\includegraphics[width=0.41\textwidth]{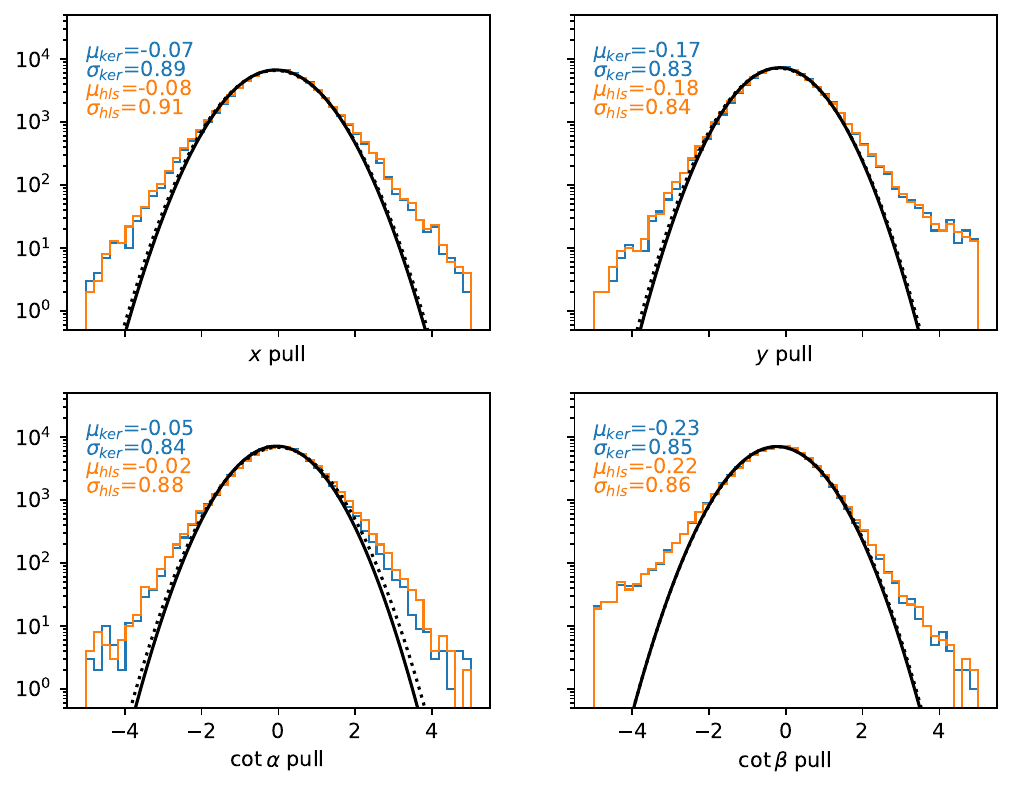}}
    \caption{A: The mean of the residual distribution and average predicted error band for QKeras (blue, cross-hatched) and \texttt{hls4ml} synthesized (orange, horizontal lines) regression networks. There are slight biases at the endpoints of the training data in the raw network output. Average predicted local coordinate errors of 5.7 microns in x, 1.1 microns in y, 3.8 degrees in the polar angle, and 1.7 degrees in the azimuthal angle are achieved. B: The distribution of the pull test-statistic for each predicted variable demonstrating that the predicted errors are 3-18\% larger than the width of the residual distributions. Excellent, but not bitwise perfect, agreement between the QKeras and \texttt{hls4ml} models is observed.}
    \label{fig:network-performance}
\end{figure}

After training the network and determining quantization parameters that achieve good agreement between the QKeras and \texttt{hls4ml} implementations of the network, results are attained as summarized in Figure~\ref{fig:network-performance}. 
To expound upon these results and demonstrate their utility, we shall focus on the predictions of angles.
Using the average predicted errors from the network on the test data sample of 600,000 clusters and using a tolerance of a factor of four on the error, the average size of a solid angle patch with high efficiency for a match is smaller than 0.02 steradians, 0.2\% of the solid angle, or a 6 cm$^2$ search region for hits in a subsequent pixel layer 4 cm away.
This implies that with such a `singlet' track seed, we can retain high track seeding efficiency without any constraints on a charged track's transverse momentum while also reducing the initial combinatorics by orders of magnitude.
Similarly, this relaxes the design requirements for track triggers that include pixel data by significantly reducing the number of initial matches to search through when, for instance, matching tracks from an outer tracker to pixel hits. In addition, this reduces needs for data replication and sharing across multiple trigger regions in the same layer.

Concerning technical aspects, it is important to note that while bit-wise equivalence is not attained with these parameters, the performance of the two implementations of the network is nearly equivalent.  There is also room for additional optimization of the QKeras implementation, particularly the bit-widths.
When synthesized into firmware with the Xilinx Vivado toolkit for the Alveo U250 accelerator card, this network requires no digital signal processors (DSP) or heavy processing resources to implement and so stands as a promising candidate for synthesis into the digital part of a front-end ASIC in a future pixel detector as seen in Table~\ref{tab:synthesis_results}.
Justifying that, the ASIC synthesis is achieved in 45~nm CMOS using Catapult HLS~\cite{catapult-hls} for a 200 MHz clock, and the logic synthesis completed successfully and can now be used to guide place and route for an ASIC design.
The differences in latency between the FPGA and ASIC syntheses can be attributed to different compiler-specific optimizations being applied, due to the different maturity of backends for Vivado and Catapult in \texttt{hls4ml}, and are likely not representative of inescapable flaws in the ASIC synthesis.
In both cases, the next major improvement in inference latency and initiation interval will come from a completely parallel implementation, as opposed to the current streaming implementation, in the \texttt{hls4ml} backends.
This implementation is not used in this work because the parallel SeparableConv2D layer is missing from \texttt{hls4ml} and is currently being implemented.
This change will remove the predominant source of latency, which is loading of the data pixel-by-pixel, though it may be replaced by additional resource usage in the spatial domain.
The next steps for this network are to fully validate its current performance in a real FPGA, and take the preliminary ASIC synthesis further including full place-and-route, understanding power usage, implementation in our target 28~nm process, and optimize input data requirements as well as implementation size.

\begin{table}[!htb]
    \centering
    \begin{tabular}{|c|c||c|c||c|c|}
        \hline
         \multicolumn{2}{|c||}{QKeras Model Analysis} & \multicolumn{2}{|c||}{Alveo U250 Synthesis} & \multicolumn{2}{|c|}{ASIC Synthesis (45nm)} \\
        \hline
         4-bit ops. & 50,140 & Clock Period & 5~ns & Clock Period & 5~ns \\
         8-bit ops. & 3,040  & BRAM\_18K & 12.5 & Area Estimate & 1.4~mm$^2$ \\
         \nth{1} Layer $N_{\mathrm{0}}$ & 17 & DSP48E & 0 & Buffer Area & 0.0017~mm$^2$ \\
         \nth{2} Layer $N_{\mathrm{0}}$  & 9 & FF & 14,289 & Inverter Area & 0.055~mm$^2$ \\
         \nth{3} Layer $N_{\mathrm{0}}$ & 2 & LUT & 57,398 & Logic Area & 0.80~mm$^2$ \\
         \nth{4} Layer $N_{\mathrm{0}}$ & 117 & URAM & 0 & Sequential Area & 0.52~mm$^2$ \\
         \nth{5} Layer $N_{\mathrm{0}}$ & 15 & Latency & 1.46~$\mu$s & Latency & 27~$\mu$s \\
         \nth{6} Layer $N_{\mathrm{0}}$ & 51 & Interval (II) & 1.38~$\mu$s & - & - \\
        \hline
    \end{tabular}
    \caption{QKeras Model Analysis (left), FPGA Firmware (middle) and ASIC (right) complete synthesis figures of merit including resource usage results. The QKeras Model Analysis provides an operation-wise breakdown of how many N-bit ops occur in the model. The number of sparse parameters, $N_{\mathrm{0}}$, per layer, is also summarized, indicating that further model compression is possible using pruning techniques. FIFO-depth optimization is included in the FPGA synthesis workflow. The ASIC synthesis is successful and represents the first attempt at using Catapult HLS for this design. The long latency and II for FPGA and ASIC syntheses are largely due to the implementation of the architecture as streaming 20 4 bit words per pixel rather than presenting all data in parallel. The amount and nature of data being fed to the model is under heavy study to achieve a realistic and low-power implementation, and this should not be taken as a fundamental limitation on the model's performance or suitability for implementation.}
    \label{tab:synthesis_results}
\end{table}

%% file: sections/5-conclusions/conclusions.tex
\section{Conclusions}

In this work we have demonstrated a highly compressed, quantized, and well performing regression neural network for predicting track states from a possible future silicon pixel sensor that is implementable in modern FPGAs and silicon lithography processes. 
This network is inspired by and improves upon techniques used in modern tracking detector pattern recognition research.
In addition to mean positions and angles, the covariance matrix for these values is predicted using a mixture density network technique, which provides the novel capability to predict efficient search windows for additional hits with a single silicon sensor.
The first estimate of the average size of this window indicates that a reduction of two orders of magnitude in search space for track seeds is possible without making assumptions on the track momentum parameters, which indicates that this network could be used as part of the the readout scheme directly on the pixel sensor.
Given the general nature of the technique used here, this sort of on-device, real time, data-compressing feature extraction with error prediction could be developed for a variety of sensing hardware that can be deployed in high data-rate experiments across the physical sciences.
The next steps of research are to start investigating network power usage and implementation as an ASIC in 28~nm CMOS, and to verify the network performance in hardware.

%% file: sections/6-acknowledgements/acknowledgements.tex
\section*{Appendix: Acknowledgements}

This work was completed using computing resources at the Fermilab Elastic Analysis Facility (EAF). We thank Burt Holzman for computing support.

We acknowledge the Fast Machine Learning collective as an open community of multi-domain experts and collaborators. This community, Javier Duarte and Vladimir Loncar in particular, were important for the development of this project.

We would like to extend our sincere gratitude to Harish Jamakhandi and David Burnette from Siemens EDA for their assistance and expertise with Catapult HLS.

DB, JD, GDG, FF, LG, JH, RL, BP, GP, CS and NT are supported by Fermi Research Alliance, LLC under Contract No. DE-AC02-07CH11359 with the Department of Energy (DOE), Office of Science, Office of High Energy Physics. JD, FF, GDG, BP, GP, and NT are also supported by the DOE Early Career Research Program. NT is also supported by the DOE Office of Science, Office of Advanced Scientific Computing Research under the “Real-time Data Reduction Codesign at the Extreme Edge for Science” Project (DE-FOA-0002501). 

AB is supported through NSF-PHY award 2013007. MS is supported by NSF-PHY award 2012584. CM is supported by NSF-PHY award 2208803.  KD is supported in part by the Neubauer Family Foundation Program for Assistant Professors and the University of Chicago. RK is supported by the Metcalf Fellowship program of the University of Chicago. AY and SK are supported by the DOE Office of Science Research Program for Microelectronics Codesign (sponsored by ASCR, BES, HEP, NP, and FES) through the Abisko Project. MSN is supported through NSF cooperative agreement OAC-2117997, the DOE Office of Science, Office of High Energy Physics, under Contract No. DE-SC0023365, REFERENCES 24 and the Discovery Partners Institute under the ``Democratizing AI Hardware with an Open-Source AI-Chip Design Toolkit'' Project.